\documentclass[12pt,a4paper,final]{iopart}

\usepackage{iopams}
\usepackage{graphicx}
\usepackage[breaklinks=true,colorlinks=true,linkcolor=blue,urlcolor=blue,citecolor=blue]{hyperref}

\newcommand{\euler}[1]{{\usefont{U}{eur}{m}{n}#1}}

\newcommand{\um}{\mbox{\euler{\char22}m }}
\newcommand{\umns}{\mbox{\euler{\char22}m}}
\newcommand{\umsqu}{\mbox{\euler{\char22}m$^2$ }}
\begin{document}

\title{Mechanical properties of branched actin filaments}

\author{Mohammadhosein Razbin$^{1}$, Martin Falcke$^{2}$,\\ Panayotis
  Benetatos$^{3}$ and Annette Zippelius$^{1}$ }

\address{$^{1}$ Max Planck Institute for Dynamics \& Selforganization, Am Fassberg 17 and Institute for
  Theoretical Physics, Georg August University, Friedrich-Hund-Platz 1, 37077 G{\"o}ttingen, Germany;\\$^{2}$Max Delbr{\"u}ck Center for Molecular Medicine, Robert R{\"o}ssle Str. 10, 13092 Berlin, and Dept. of Physics, Humboldt University, Newtonstr. 15, 12489 Berlin, Germany;\\$^{3}$ Department of Physics, Kyungpook National University, 80 Daehakro, Bukgu, Daegu 702-701, Korea}

\ead{m.razbin@theorie.physik.uni-goettingen.de ~ pben@knu.ac.kr}

\begin{abstract} 
  {Cells moving on a 2dimensional substrate generate motion by polymerizing actin filament networks inside a flat membrane protrusion. New filaments are generated by branching off existing ones, giving rise to branched network structures. We investigate the force-extension relation of branched filaments, grafted on an elastic structure at one end and pushing with the free ends against the leading edge cell membrane. Single filaments are modeled as worm-like chains, whose thermal bending fluctuations are restricted by the leading edge cell membrane, resulting in an effective force. Branching can increase the stiffness considerably; however the effect depends on branch point position and filament orientation, being most pronounced for intermediate tilt angles and intermediate branch point positions. We describe filament networks without cross-linkers to focus on the effect of branching. We use randomly positioned branch points, as generated in the process of treadmilling, and orientation distributions as measured in lamellipodia. These networks reproduce both the weak and strong force response of lamellipodia as measured in force-velocity experiments. We compare properties of branched and unbranched networks. The ratio of the network average of the force per branched filament to the average force per unbranched filament depends on the orientation distribution of the filaments. The ratio exhibits compression dependence and may go up to about 4.5 in networks with a narrow orientation distribution. With orientation distributions measured in lamellipodia, it is about 2 and essentially independent from network compression, graft elasticity and filament persistence length.
}

\end{abstract}

\vspace{2pc}
\\{Accepted in Physical Biology}

\section{Introduction}

The crawling of many different cell types is essential for life. Undifferentiated cells move towards the site, where they form a tissue or organ in the developing embryo. Skin cells start crawling when they have to close a wound~\cite{bray2001}. During metastasis, cancer cells dissociate from the primary tumor, crawl towards blood vessels and spread all over the body~\cite{Yamaguchi2005,Condeelis2006}. Branched actin filaments carry forces during cell motion, and consequently understanding their elastic properties is central to understanding the mechanics of cell motility.

In vitro, cells are plated on a two dimensional substrate to observe their dynamics. They form a flat membrane protrusion in the direction of motion, the lamellipodium, which is only about 100-200~nm thick but several $\um$ deep and wide~\cite{Small2002}. A dense network of actin filaments (F-actin) inside the lamellipodium pushes the leading edge membrane forward~\cite{Svitkina97}. Treadmilling of the filaments drives motion~\cite{Pollard2003}: Filament barbed (or plus) ends polymerize at the leading edge of the lamellipodium and the pointed (or minus) ends depolymerize at the rear.

Usually cells move in response to an external signal. A variety of signals stimulate the activation of nucleation promoting factors (NPFs) (like WASp or WAVE) located in the leading edge membrane of the lamellipodium. They activate the actin related protein complex Arp2/3. It binds to an existing filament very close to or at its barbed end at the lamellipodium's leading edge. That initiates the growth of a new filament branch out of the Arp2/3 complex. Many of these branched structures consisting of mother filament and branch form the F-actin network in the lamellipodium. The branched structure itself is dynamic. The branch point with the Arp2/3 complex moves rearward due to treadmilling in the same degree as mother filament and branch grow. Since Arp2/3 binding to the individual filaments is not synchronous we find at any time many different positions of branch points in the lamellipodium F-actin network.

The elastic properties of the F-actin network crucially depend on the density of links between filaments~\cite{pollard98b,Mogilner2003,gardel2004,Bohnet2006,Radmacher2006,radmacher2011,falcke2012_BPJ}. Molecular links arise in two ways: Cross-links connecting two filaments at some point along their contour length are formed by cross-linker molecules like filamin or $\alpha$-actinin, and branching attaches the minus end of a filament laterally to a mother filament. Intuition suggests that branching alone could stiffen the network to some degree, since branching is a geometrical constraint on the configuration of two filaments. That intuition has never been quantified before, but is supported by our results presented below. On the other hand, the network region close to the leading edge was found to be as soft as \emph{weakly} cross-linked actin networks~\cite{gardel2004,Bohnet2006,Radmacher2006,radmacher2011,falcke2012_BPJ}, and experiments in actin solutions suggest that branching contributes very little to the elastic modulus of F-actin networks~\cite{stossel2002}. Here, we would like to present a first step in quantifying the contribution of branching to the elastic and semi-flexible properties of the lamellipodial F-actin network. How much stiffer than single filaments are branched filaments? How are their properties reflected in network behavior? We will answer these questions by investigating a single branched filament and networks of branched structures in an approximation neglecting interactions between them in order to focus on branching effects. This neglect of interactions implies that we consider only elastic properties on short time scale and not the visco-elastic properties arising from cross-linking.

The mother filament is grafted at one end and has a free tip at the other one in our model system (see Fig.~\ref{fig:configuration}). The graft is provided by a highly cross-linked part of the F-actin network. This idea is based on the increasing degree of cross-linking and filament bundling towards the rear of the lamellipodium, which has been observed in many different experiments and simulations~\cite{Svitkina97,svitkina99,borisy99a,Danuser2005a,theriot2008,falcke2010_PRE,mogilner2009a,falcke2010_BPJ,mogilner2011,levine2012} (see~\cite{falcke2014_EPJST} for a detailed discussion). The graft moves in the direction of cell motion due to cross-linker binding and bundling, and thus contour length 'flows' into the graft. In the steadily moving cell, the balance between polymerization and cross-linking creates a stationary distance between graft and leading edge membrane. At any time, the branched structure is in a configuration similar to Fig.~\ref{fig:configuration}, but the branch point moves with treadmilling towards the graft point. Both mother filament and branch polymerize such that their barbed ends stay at the leading edge membrane. Consequently, we will vary the position of the branch point on the mother filament from 0 to its full length $L$ when we investigate the elastic properties. We model the filaments in the weakly bending regime, i.e. bending does not affect the end-to-end distance. We allow for an elastic graft of stiffness $K_s$ and model the membrane by a constraint, enforcing the filaments to be entirely on the left side of the membrane.

After the introduction of the model, we will discuss results for a single branched filament. The F-actin network in the lamellipodium comprises filaments with many different tilt angles and branch point positions. Hence, we consider it as an ensemble of branched structures with varying branch point and tilt angle and calculate its properties as the average over this ensemble. The network average of the force will be compared to experimental results from force-velocity measurements and the force of unbranched networks.

\section{Results}
\subsection{The model: semiflexible branched filaments in the weakly bending limit}
We model the interior of the lamellipodium as a two-dimensional space, since it is approximately flat as described above. Furthermore, the results are easily generalised to three dimensions. We always assume a sufficiently large persistence length, such that the weakly bending approximation applies and the fluctuations perpendicular to the mean orientation of the polymer segment are small and can be treated on a Gaussian level.

Our elemental structure is a grafted filament of contour length $L$. The probability to find its tip at position $(x,y)$ with orientation $\theta$, given that it is grafted at $(x_0,y_0)$ with orientation $\omega$ is denoted by $G_L(x,y,\theta|(x_0,y_0,\omega)$. For the simple case of perpendicular grafting, $\omega=0$, at $(x_0,y_0)=(0,0)$, $G_L$ obeys \cite{Benetatos03}
\begin{equation}
  G_{L}(x,y,\theta|0,0,0) \propto \mbox{exp}[-\frac{3l_{p}}{L^{3}}(y^{2}-Ly\theta+\frac{L^{2}\theta^{2}}{3})]\delta (x-L).
\label{eq:Green1}
\end{equation}
The general case is obtained by a translation and rotation according to
\begin{eqnarray}
\theta&\to& \theta-\omega\nonumber\\
y & \to & (y-y_0)\cos{\omega}-(x-x_0)\sin{\omega}\nonumber\\
x & \to & (x-x_0)\cos{\omega}+(y-y_0)\sin{\omega}
\end{eqnarray}
and explicitly given by Eq.~\ref{eq:GL_tilted}.

\begin{figure}[t]
\centering{\includegraphics[width=0.45\textwidth]{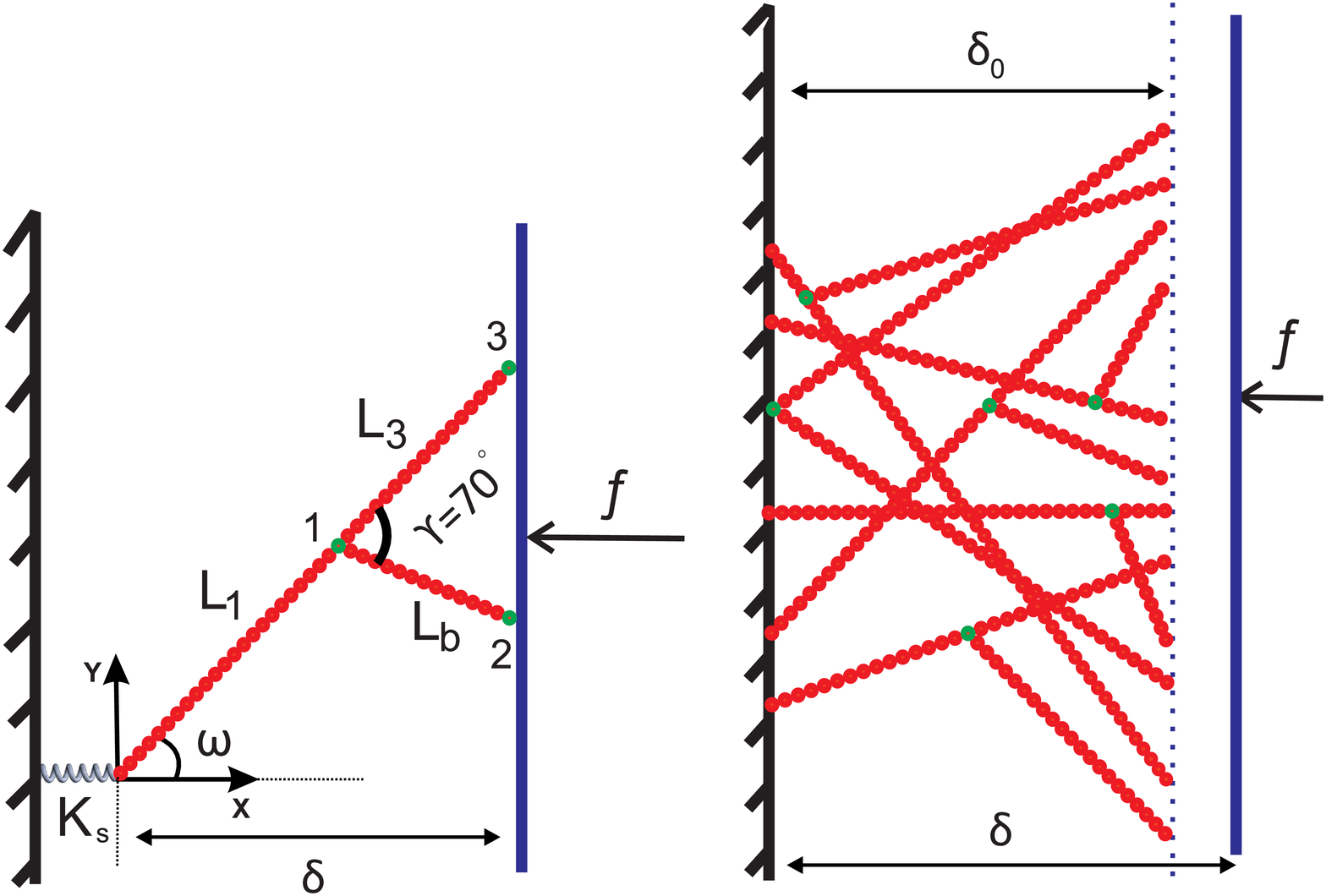}}
\caption{Left: Single branched filament grafted on a soft graft and confined in the x-direction by a flat membrane. Thermal bending fluctuations are not shown in this drawing. $L_1$ is the contour length between graft point and branch point, $L_3$ is the contour length between branch point and filament tip. The contour length of the mother filament is $L=L_1+L_3$, $L_b$ is the branch contour length. The branch angle is $\gamma=70^{\circ}$ throughout the study, and $\omega$ is the tilt angle. The numbers 1, 2 and 3 refer to the filament tip coordinates in Eq.~\ref{branched1}. Right: Network of branched filaments with various orientations and locations of branch points. The length $\delta_0$ is the distance of the filament tips from the graft plane without bending and fluctuations. The lengths $L$ and $L_b$ obey $L=\frac{\delta_0}{\cos(\omega)}, L_b=\frac{L_3\cos(\omega)}{\cos(\omega-\gamma)}$, and $\delta$ denotes the distance between the graft point and the leading edge membrane. }
\label{fig:configuration}
\end{figure}

In a first step we compute the probability, $P_{t }(x|x_0)$, that the endpoint of the tilted polymer is at a distance of $x-x_0$ from the graft point:
\begin{equation}
P_{t }(x|x_0)\equiv \int \int dyd\theta\  G_{L}(x,y,\theta|x_0,0,\omega).
\end{equation}
We consider an elastic, fluctuating structure into which the filament is grafted and hence model it by a fluctuating spring in x-direction with spring constant $K_s=(k_BT)K$, zero equilibrium length and a distribution of the spring extension proportional to $\exp(-(Kx_0^2)/2)$.
The stiffness of the substrate, $K$, is assumed to be large as compared to the stiffness for transverse fluctuations, $K\gg K_{\perp}=\frac{2 l_p}{3L^3}$, of a filament of contour length $L$. In this study we take $K=100\frac{2 l_p}{3(L\cos(\omega))^3}$. The probability $P_{t }(x)$ to find the filament tip at $x$ then follows by averaging over all graft point positions $x_0$:
\begin{equation}
P_{t }(x)= \int \frac{dx_0}{\sqrt{2\pi K}}\; P_t(x|x_0)\exp(-\frac{Kx_0^2}{2}).
\label{endpoint1}
\end{equation}
The filament exerts a force on an impenetrable flat membrane at a distance $\delta$ from the graft plane (see Fig.~\ref{fig:configuration}). The force originates from the reduction in the number of filament configurations due to the constraint $x-x_0\leq \delta$ imposed by the membrane; we therefore call it an entropic force. The fraction of configurations satisfying the constraint is given by
\begin{equation}\label{partition_function}
Z_{t}(\delta)=\int dx P_{t}(x) \Theta (\delta-x).
\end{equation}
The derivative of $Z_t$ with respect to $\delta$ is the entropic force~\cite{Gholami2006}
\begin{equation}\label{entropic}
f_{t}(L,\delta,\omega)=k_B T \frac{\partial}{\partial \delta} \ln Z_{t}(\delta).
\end{equation}

These results for a single grafted polymer are easily generalised to the branched structure shown in Fig.~\ref{fig:configuration}. The probability to find the two endpoints at $x_2$ and $x_3$ respectively given that the structure is grafted at point $x_0$ is calculated by the following expression:
\begin{eqnarray}\nonumber
\lefteqn{P_b(x_{2},x_{3}\mid x_0)=}&&\\
&&\int \int \int dx_1dy_1d\theta_1 G_{L_1}(x_1,y_1,\theta_1 \mid x_0,0,\omega)\nonumber\\
&&\times \int \int dy_2d\theta_2 G_{L_b} (x_2,y_2,\theta_2 \mid x_1,y_1,\theta_1-\gamma)\nonumber\\
&&\times \int \int dy_3d\theta_3 G_{L_3} (x_3,y_3,\theta_3 \mid x_1,y_1,\theta_1)
\label{branched1}
\end{eqnarray}
The stiffness of the graft is again taken into account by averaging over the positions of the graft:
\begin{equation}
P_{b }(x_2,x_3)\propto \int dx_0 P_b(x_2,x_3|x_0)\exp(-\frac{Kx_0^2}{2})
\end{equation}
The explicit expression for $P_b(x_2,x_3)$ is given by Eq.~\ref{eq:Pb} in the Appendix. The fraction of configurations satisfying the constraint obeys in analogy to Eq.~\ref{partition_function}
\begin{equation}
Z_b(\delta)=\int dx_2\int dx_3 P_b(x_2,x_3) \Theta (\delta-x_2)\Theta (\delta-x_3),
\end{equation}
and the entropic force follows from
\begin{equation}\label{entropic}
f_b(L,L_b,L_3,\delta,\omega)=k_B T \frac{\partial}{\partial \delta} \ln Z_b(L,L_b,L_3,\delta,\omega).
\end{equation}
The force $f_b$  depends on the parameters $L\equiv (L_1+L_3)$, $L_3$, $\omega$ and $\delta$ ($L_b$ is fixed by $\omega$, $L$ and $L_3$).
We will present our results as dimensionless quantities, and scale lengths by $\delta_0$ and force by $k_BT/\delta_0$ for that purpose. We comment on which specific values of parameters are suggested by experimental observations in the Discussion and the Appendix below.

\subsection{Properties of single branched filaments}

\begin{figure}[t]
\centering{\includegraphics[width=0.6\textwidth]{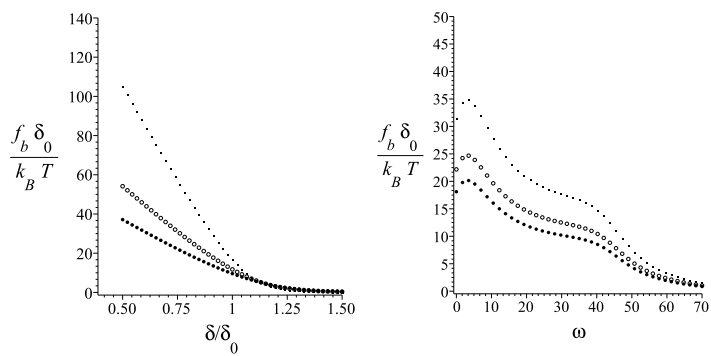}}
\caption{Left panel: force versus relative distance, $\delta/\delta_0$, between membrane and graft point for the symmetric case $\omega=35^{\circ}$ and $L_3=L_b=\frac{L}{2}$. Right panel: force versus tilt angle $\omega$ for $\delta=\delta_0$. The curves represent the ratio of $\frac{l_p}{\delta_0}=10,\; 5$ and $\frac{10}{3}$ from top to bottom in both figures.}
\label{fig:Branchedactin}
\end{figure}

The entropic force exerted by the branched filament on the membrane is shown in the left panel of Fig.~\ref{fig:Branchedactin} for the symmetric case $\omega\sim 35^{\circ}$ and branching at the midpoint of the mother filament. The force decreases with increasing distance $\delta$ between grafting plane and membrane. An infinitely stiff filament ($l_p$ very large) would just touch the membrane, if $\delta$ equals $\delta_0=L\cos{\omega}$. But the force exerted by a semiflexible filament is nonzero even for $\delta>\delta_0$, because the tilted branched structure exhibits fluctuations with the endpoint reaching beyond $\delta_0$. In this regime the force is nearly independent of persistence length, whereas for $\delta<\delta_0$ we observe a strong increase with $l_p$.

\begin{figure}[h]
\centering{\includegraphics[width=0.6\textwidth]{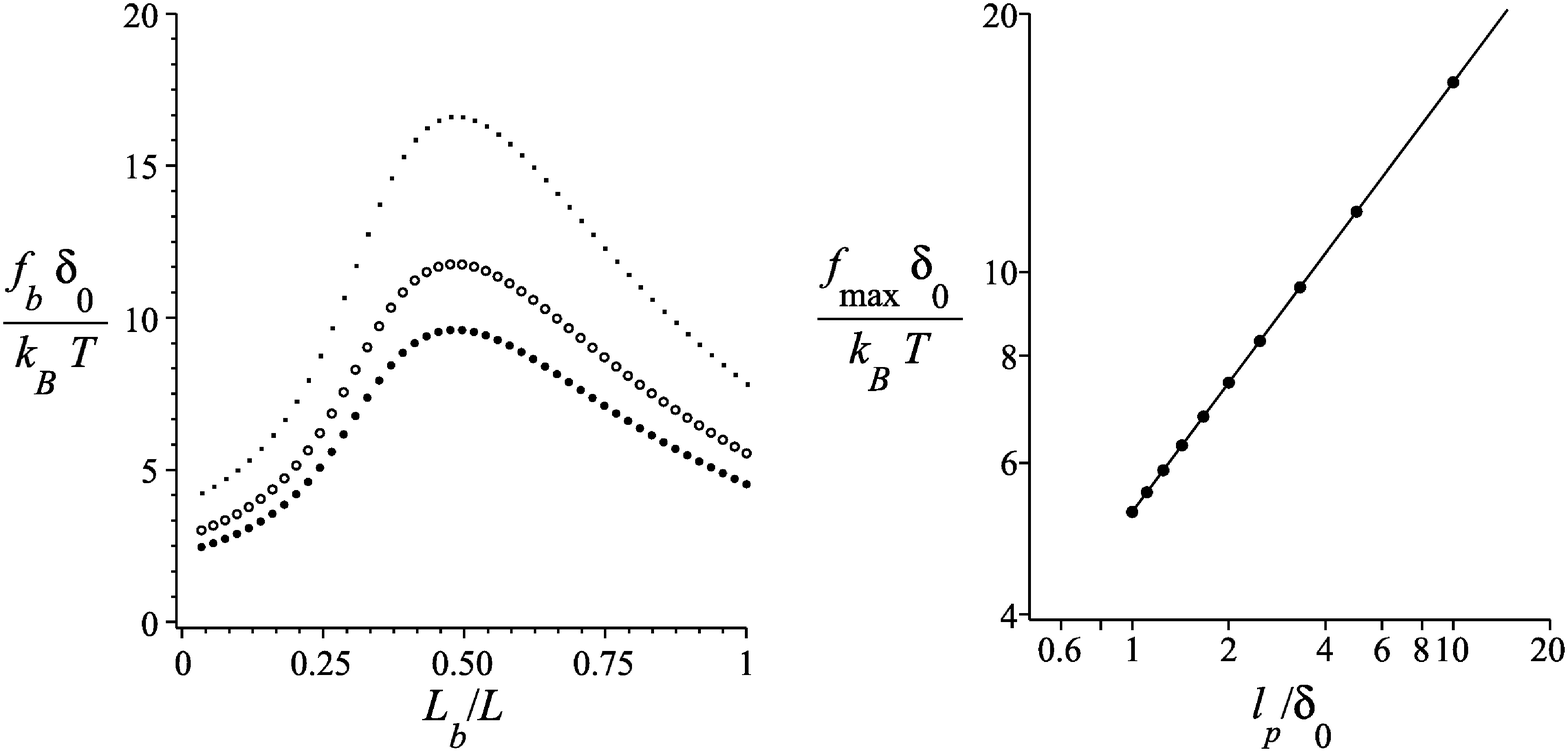}
\includegraphics[width=0.4\textwidth]{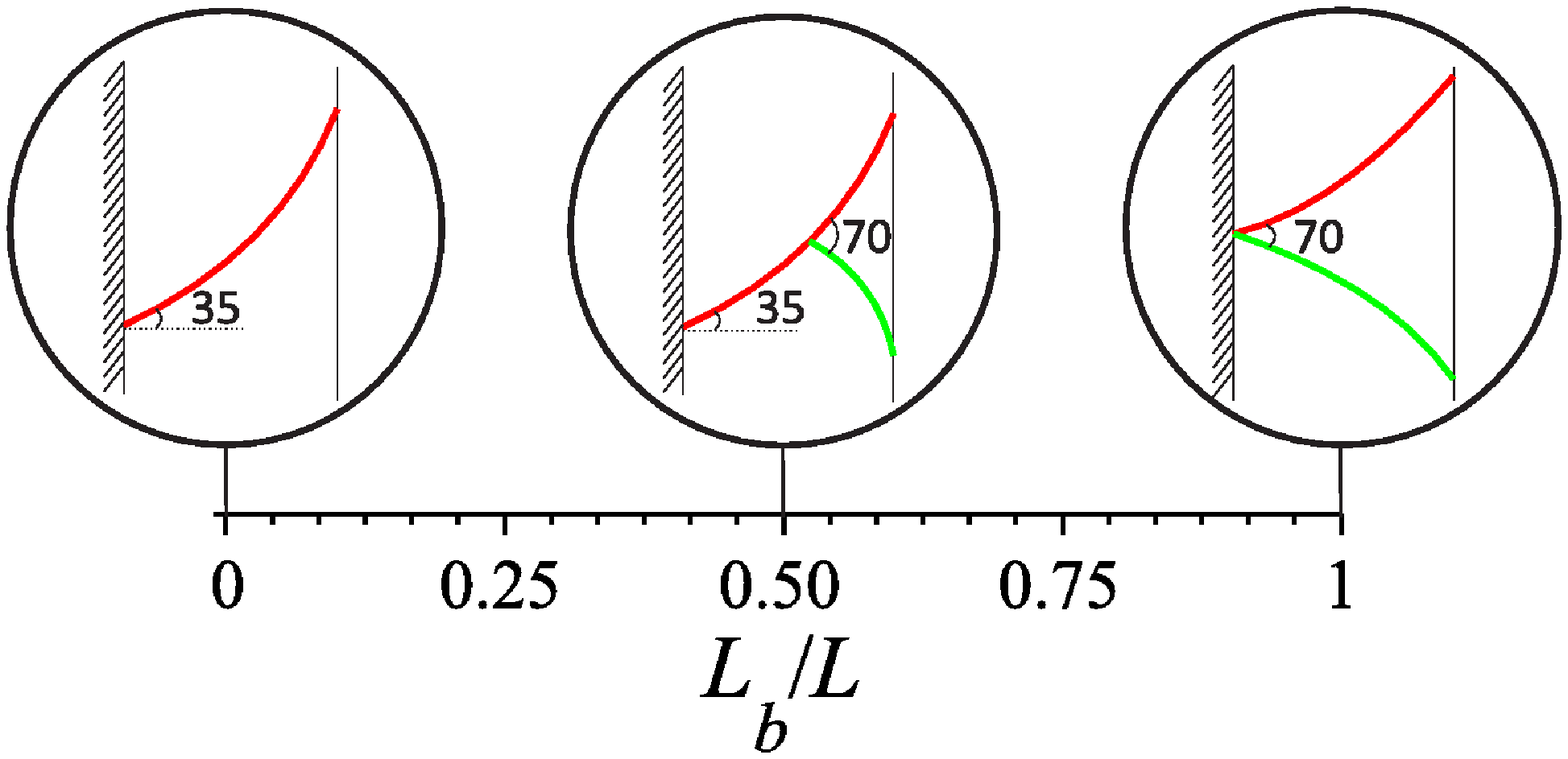}}
\caption{Left: force versus relative length of the daughter branch. Below: the configurations of the branched filament corresponding to $L_b/L=0, 0.5, 1.0$. The parameter values are: $ 
  L_3=L_b,\; \omega=35^{\circ}$, $\delta=\delta_0$. From high to low force values the symbols represent the ratio of $\frac{l_p}{\delta_0}=10,\; 5$ and $\frac{10}{3}$, respectively. Right: log-log plot of the maximum force as a function of persistence length $l_p$, fitted to a square root dependence $f_{max}\propto \sqrt{l_p}$. }
\label{fig:graftpoint}
\end{figure}

\begin{figure}[h]
\centering{\includegraphics[width=0.6\textwidth]{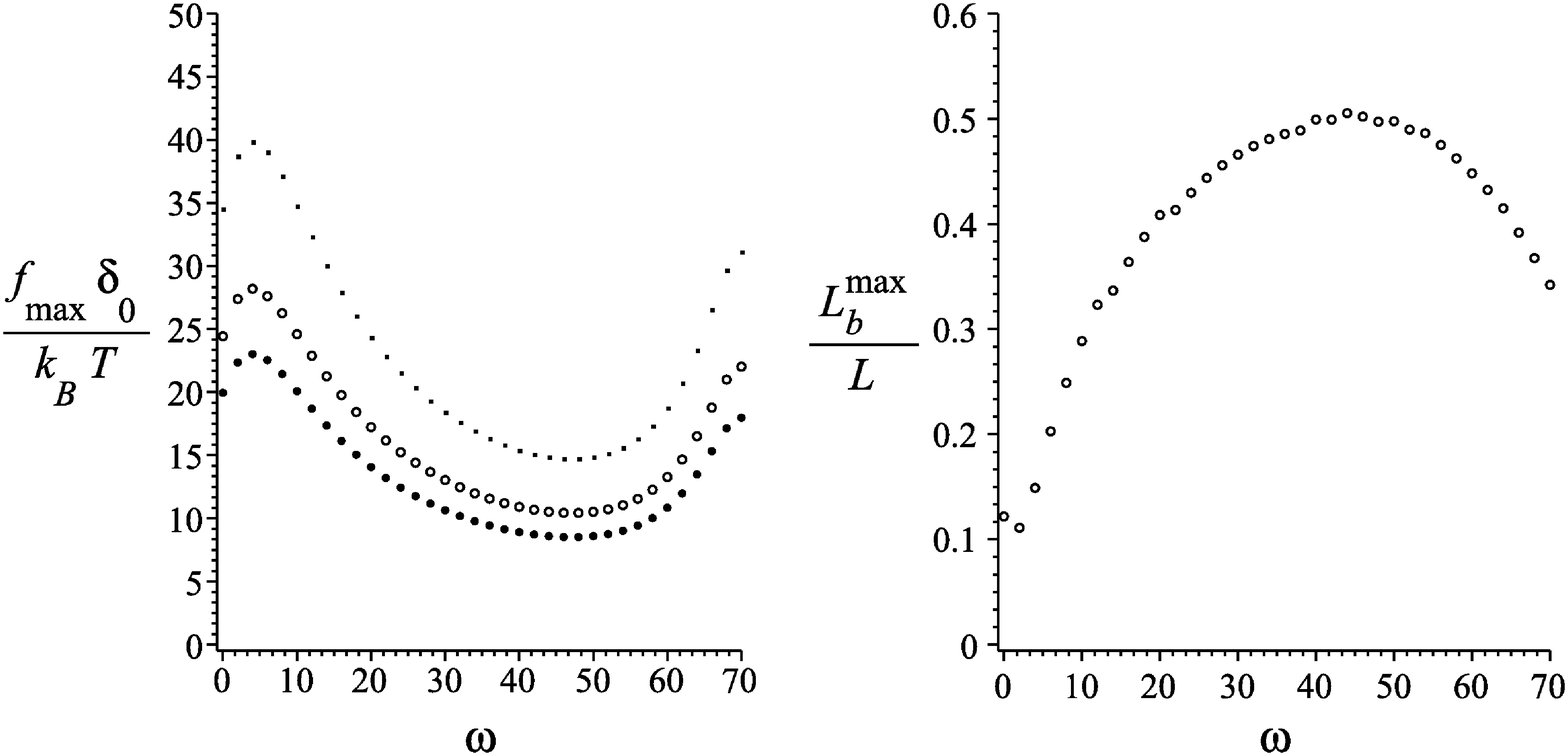}
\includegraphics[width=0.6\textwidth]{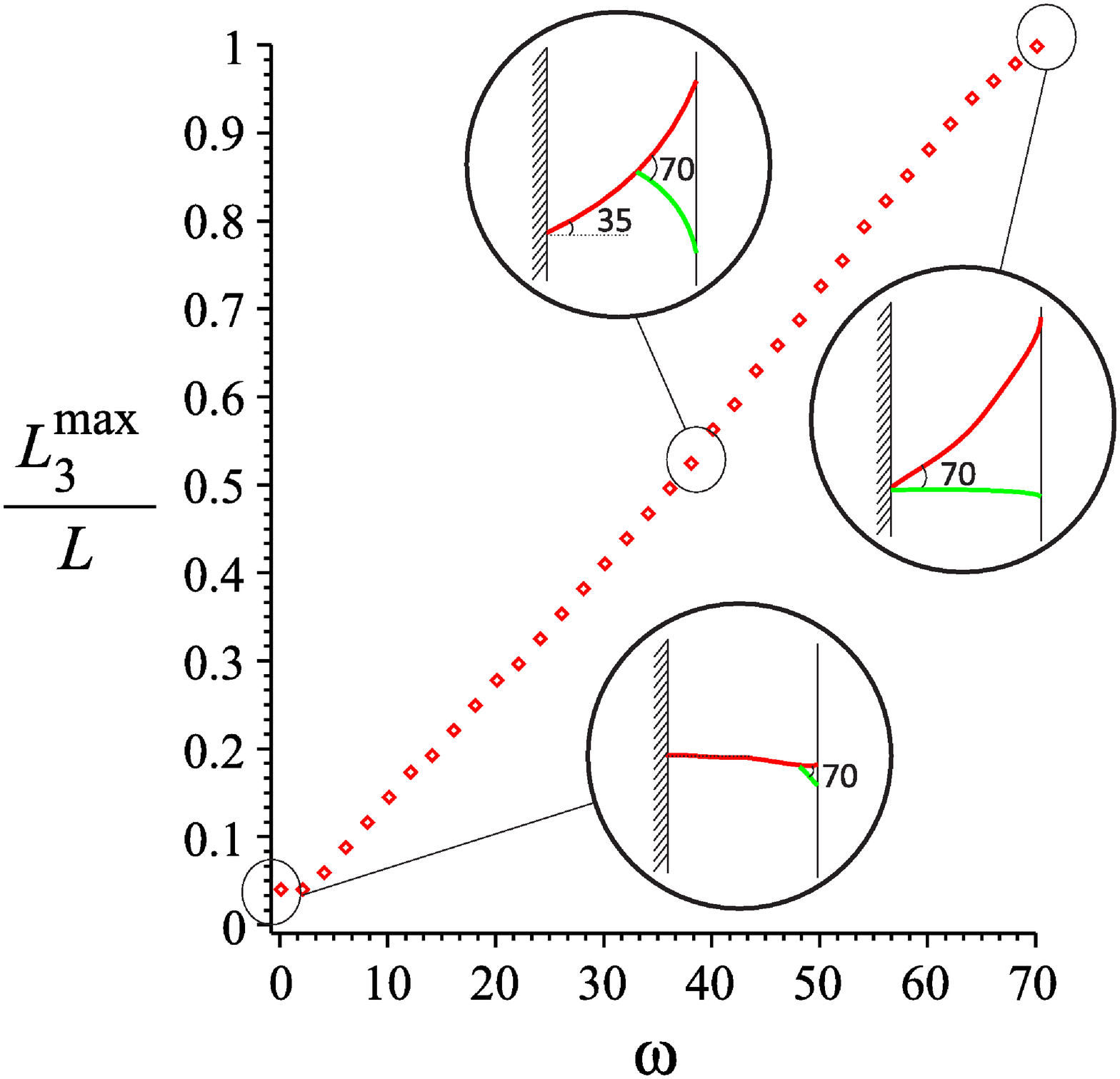}}
\caption{Upper left: Maximum of the force $f_{max}$ obtained by varying $L_b$, versus tilt angle $\omega$; we have $\delta=\delta_0 $ for the curves. From high to low force values the curves represent the ratio of $\frac{l_p}{\delta_0}=10,\; 5$ and $\frac{10}{3}$, respectively. Upper right: Relative branch length $L_b^{max}/L$ of the maximum force versus tilt angle $\omega$ for the same parameters (independent of the persistence length $l_p$). Bottom: Relative partial contour length of the mother filament in between the endpoint and the branch point with maximal force $L_3^{max}=L_b^{max}\cos(\omega-\gamma)/\cos \omega$; also shown are three representative configurations.}
\label{fig:graftpointomega}
\end{figure}

The force exerted by the filament on the membrane is crucially affected by the tilt angle. It is plotted as a function of tilt angle $\omega$ for fixed $\delta=\delta_0$ in the right panel of Fig.~\ref{fig:Branchedactin}. Remarkably, there is a shoulder in all three curves around $\omega\sim 35^{\circ}$, indicating that the symmetric structure generates comparatively large forces. For $\omega\to 0$, the force would diverge for a stiff graft, since Eq.~\ref{eq:Green1} excludes longitudinal fluctuations of the filament tip. The high but finite stiffness of the graft limits the force which can be exerted on the membrane. These two opposite effects generate the maximum in the force close to $\omega=0$.

Figure~\ref{fig:graftpoint} shows the entropic force of a branched structure for the whole range of branch point positions on the mother filament. At the time of Arp2/3 binding to the mother filament ($L_b$=0), the complete structure has of course the properties of the mother filament. When the branch point has reached the graft plane, the branched structure corresponds to two filaments with the corresponding tilt angles. In between, the force has a maximum at $L_b=0.484L$. The maximum force is about 2.25 times the force of two filaments with length $L$. For the special case under consideration, $\delta=\delta_0$, the maximum of the force scales like the square root of $l_p$, as shown in the right panel of Fig.~\ref{fig:graftpoint}. For the compressed case, $\delta<\delta_0$, we observe a crossover to linear scaling. 
To understand this behaviour of the force, we consider the case of a single filament: The force has a thermal (entropic) and a nonthermal (energetic)
contribution. The thermal force dominates for $\delta\geqslant\delta_0$. Evaluating the force given in Eq~\ref{eq:force of single filament} 
for $\delta=\delta_0$ and large $K$, we explicitly see $f\propto \sqrt{l_p}$. As we compress the filament the energetic force 
becomes the dominant contribution. It can be computed as the force to bend a grafted cantilever beam 
resulting in $f=(\frac{3l_pk_BT}{2L^3})\frac{(\delta-L\cos(\omega))}{\sin^2(\omega)}$. 
Hence we observe a crossover from the scaling of the force with $\sqrt{l_p}$ to linear scaling as $\delta$ is decreased below $\delta_0$

Both, the maximum force, $f_{max}$, as well as its branch point position, $L_b^{max}$, depend on the tilt angle $\omega$. The general dependence on $\omega$ is shown in the top left panel of Fig.~\ref{fig:graftpointomega} for $f_{max}$ and in the top right panel for $L_b^{max}$. Since the dependencies are nonmonotonic, we plot in the bottom panel the configurations which give rise to the maximum force and help to understand the non-monotonic dependence. For very small $\omega$, i.e. almost perpendicular grafting, $L_b^{max}$ is very small and hence also the distance between the branch point and the endpoint of the mother filament, denoted by $L_3^{max}$. As the tilt angle, $\omega$, increases, the branch point moves further away from the leading edge membrane and hence both, $L_b^{max}$ and $L_3^{max}$ increase. As $\omega$ approaches $70^{\circ}$, the branch is almost perpendicular to the membrane and the maximum force is obtained for branching at the grafting plane, implying $L_3^{max}=L$. Plotting $L_b^{max}/L$ as in the top right panel of Fig.~\ref{fig:graftpointomega}, one actually observes a decrease of $L_b^{max}/L$, because $L$ grows faster than $L_b^{max}$ as $\omega \to 70^{\circ}$.

$L_3^{max}$ is independent of the persistence length, but does depend on $\delta$. For $\delta\ge\delta_0$ entropic contributions dominate and the force is largest for two independently fluctuating filaments, such that $L_3^{max}/L$ quickly approaches 1 as $\delta$ extends beyond $\delta_0$. On the other hand, for $\delta<\delta_0$, elastic
contributions are important. In the symmetric case, we find $0.435\le L_3^{max}/L\le 0.484$ for $0.5\le\delta/\delta_0\le1.0$.

The maximum force is observed for $\omega\sim 0$ and $\omega\sim 70^{\circ}$, because either the mother filament or the daughter filament are perpendicular to the membrane (upper left panel of ) . When the branch is perpendicular, the force is mainly determined by the fluctuations of the branch point, and when the mother filament is perpendicular by the fluctuations of the graft point. Moving away from perpendicular incidence the force has to decrease, giving rise to a minimum for intermediate $\omega$.

\begin{figure}[t]
\centering{\includegraphics[width=0.45\textwidth]{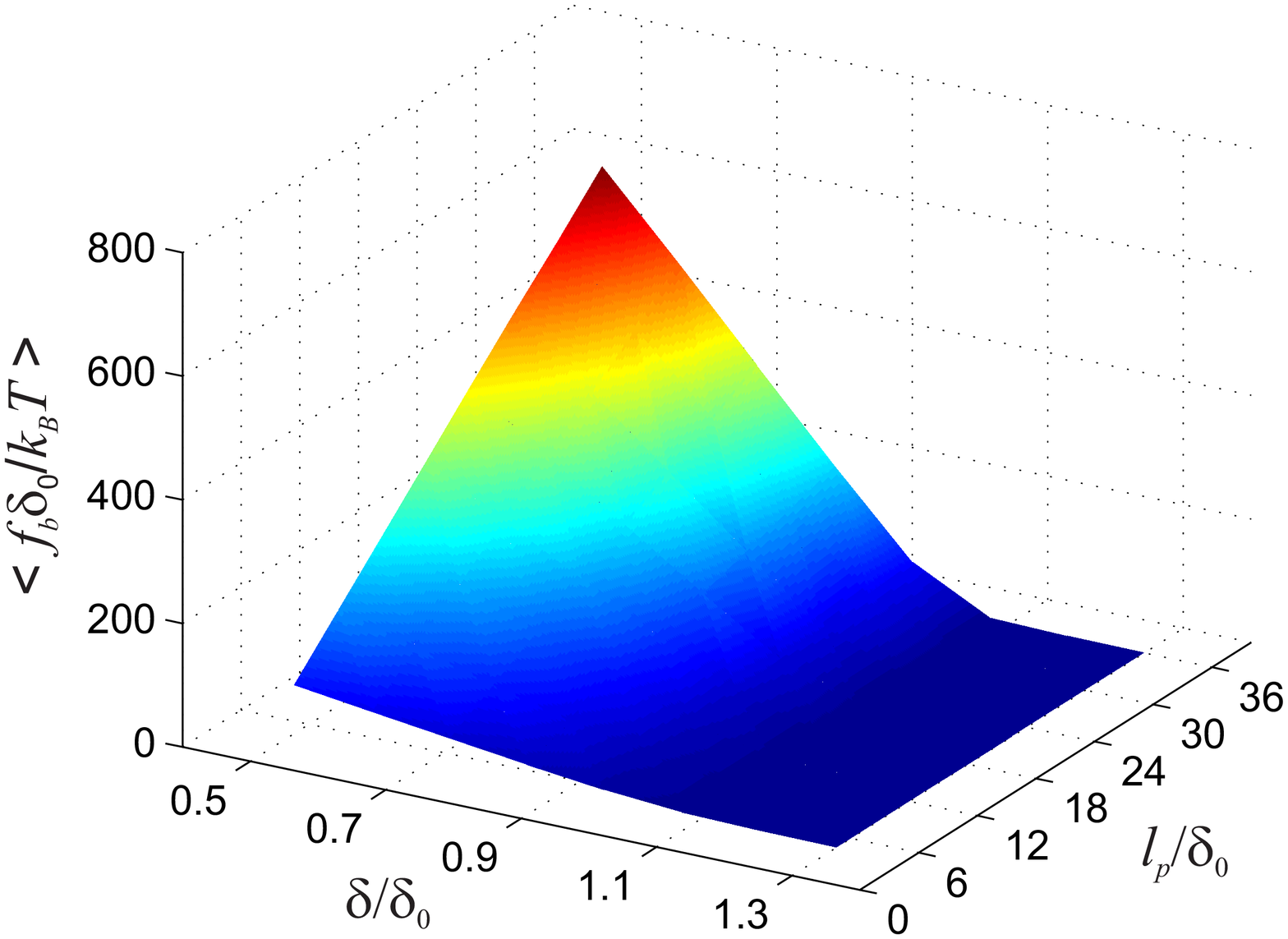}\includegraphics[width=0.45\textwidth]{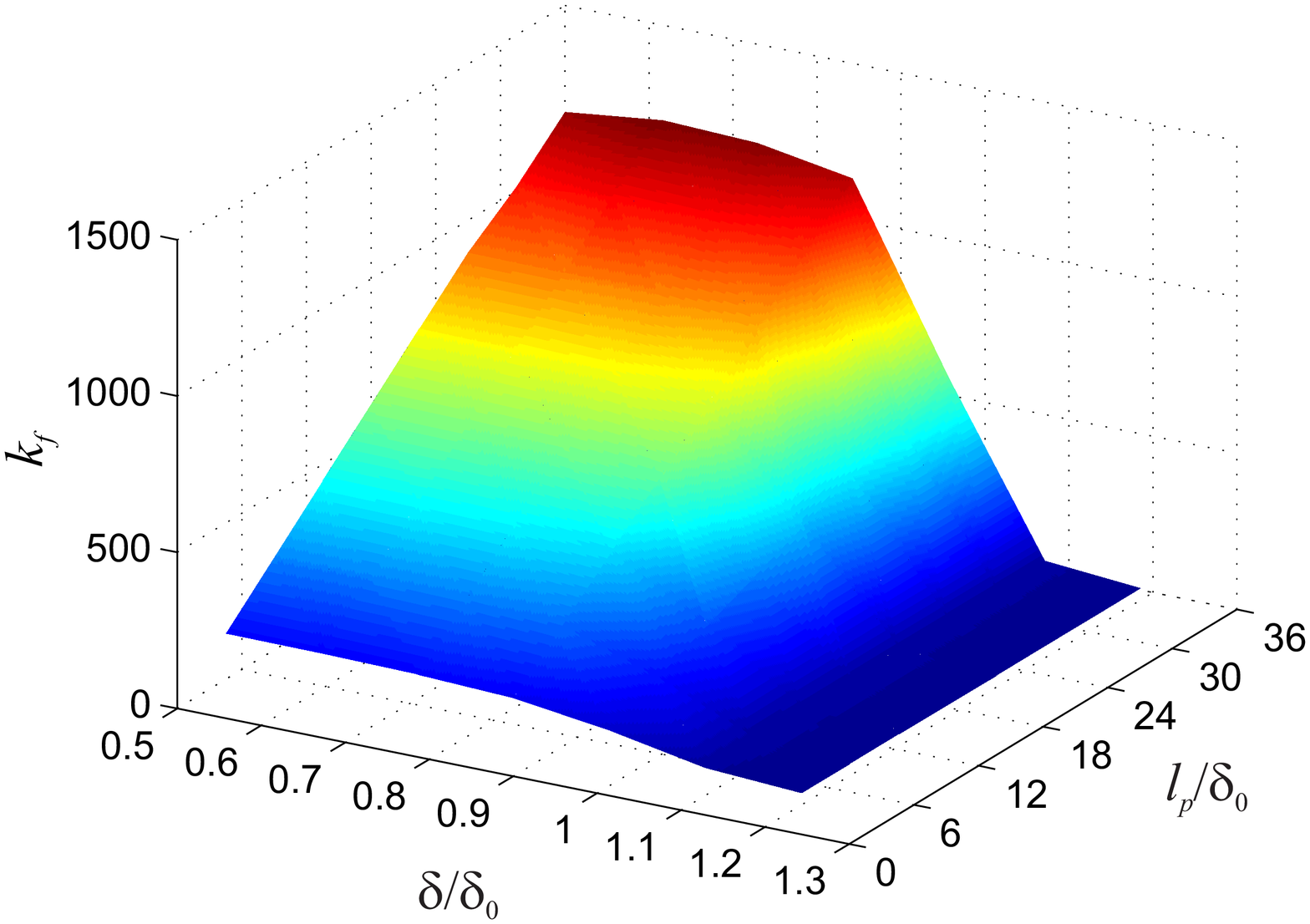}}
\caption{Left panel: Average force per branched structure of the network as a function of $\delta/\delta_0$ and $l_p/\delta_0$. Right panel: The force constant $k_f$, which is the derivative of $f_b\delta_0/k_BT$ with respect to $\delta/\delta_0$.}
\label{fig5}
\end{figure}

\begin{figure}[t]
\centering{\includegraphics[width=0.35\textwidth]{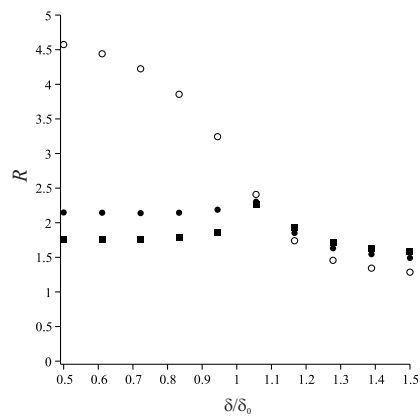}}
\caption{Ratio $R$ of the average force per filament of branched to unbranched networks versus relative distance $\delta/\delta_0$ between the membrane and the graft point; $\frac{l_p}{\delta_0}=10$. The squares refer to the random network with a uniform orientation distribution in the range $-70^\circ\leq\omega\leq70^\circ$ as measured in ref.~\cite{Small2008}, the circles show results with a narrow uniform distribution of orientations in $30^\circ \leq \omega \leq 40^\circ$,  the dots refer to the orientation distribution given by Eq.~\ref{eq:malydist} as measured in ref.~\cite{borisy2001}.}
\label{fig6}
\end{figure}

\subsection{Extension to the F-actin network: Properties of an ensemble of branched filaments}

We would like to obtain an estimate of how branching affects the network properties. With the theoretical means set up above and in the spirit of the study, we calculate the force as an average across an ensemble of branched structures, in which all interactions between the branched filaments are neglected. The ensemble is described by branch point ($L_3$)and orientation ($\omega$) distributions. The branch point is equally likely anywhere on the mother filament, corresponding to $0\leq L_3\leq\delta_0/\cos\omega$. The length of the branch obeys $L_b=\frac{L_3\cos\omega}{\cos(\omega-\gamma)}$. As far as the orientation of the mother filaments is concerned, several scenarios have been discussed in the literature. In ref.~\cite{Small2008}, electron microscopy was used to determine the orientation of filaments in lamellipodial actin networks.  The distribution was found to be approximately uniform in the range of angles between 0 and 60$^\circ$ with a small contribution between between 60$^\circ$ and 75$^\circ$. We describe it here as a uniform distribution between -70$^\circ$ and 70$^\circ$.

The force for fixed $\delta_0$ averaged over branch point positions and tilt angles is given by
\begin{eqnarray}\label{eq:average_force}
\lefteqn{<f_b(\delta)> = \int_{\omega_{min}}^{\omega_{max}} d\omega
\frac{\cos{\omega}}{\delta_0\Delta\omega}}&&\\ && \times\int_0^{\frac{\delta_0}{\cos{\omega}}} d L_3\: f_b\big(\frac{\delta_0}{\cos{\omega}},
\frac{L_3\cos\omega}{\cos(\omega-\gamma)}
,L_3,\delta,\omega\big).\nonumber
\end{eqnarray}
where $\Delta\omega=\omega_{max}-\omega_{min}$ denotes the range of the distribution. The average force is a monotonic function of both, the persistence length and the distance $\delta$ between grafting plane and membrane, as shown in Fig.~\ref{fig5}.

We can compare our results with measurements of the force-velocity relation of fish
keratocyte lamellipodia. The forces exerted by the leading edge of the freely
running cell immediately upon collision with the cantilever were below the force
resolution of the cantilevers~\cite{falcke2012_BPJ,Radmacher2006,radmacher2011}, but
caused an immediate decrease of leading edge velocity by 1-3 orders of magnitude.
Hence, the leading edge is much softer than the softest cantilever used in the
experiments, which had a force constant of 9.1 nN/\umns. If the leading edge had the
same force constant, a single branched structure would need to have a constant of
about 11.8~pN/\um with 220~filaments/\um~\cite{small2010} and a contact length of
about 7~\um~\cite{radmacher2011}. The value of $\delta_0$ in the freely running cell
was estimated to be $\sim$1.3~\um~\cite{falcke2012_BPJ}, and such a value is also
supported by the data in ref.~\cite{small2011}. Hence, if the leading edge had the
same force constant as the softest cantilever, the dimensionless force constant of a
single branched structure would be 11.8~pN\umns$^{-1}\delta_0^2$(k$_B$T)$^{-1}$=4748
(k$_B$T=4.2 10$^{-3}$~pN\umns). The values in the right panel of
Fig.~\ref{fig5} show that the branched filaments are much softer on average, which
is in agreement with the dramatic velocity drop of the lamellipodium leading edge
upon collision with the cantilever.

We do not know measurements of the pressure exerted by the filaments on
the leading edge membrane in the freely running fish keratocyte but can estimate it
from our results. Assuming $\delta_0$=1.3~\umns, $l_p/\delta_0\approx$ 10,
$\delta/\delta_0\approx$ 0.8~\cite{falcke2012_BPJ} we obtain $f\delta_0/k_BT\approx$
20 and a force per branched structure of 0.063~pN.  The pressure exerted by the
filaments on the leading edge membrane is in the range of 46 Pa (with 110 branched
structures per micrometer and a lamellipodium height of 150~nm as
in~\cite{radmacher2011}).

The value of $\delta_0$ decreases during the force-velocity measurement due to a
dynamic equilibrium between polymerization and cross-linking~\cite{falcke2012_BPJ}.
In the stalled state, $\delta_0\approx$0.27~\um applies~(Fig.~S3 of
\cite{falcke2012_BPJ}). The measured stall pressure exerted on the
leading edge by the cantilevers is 300-750~Pa~\cite{falcke2012_BPJ} and
110-430~Pa~\cite{radmacher2011}. These pressures correspond to stall forces of
0.045-0.1125~nN~\cite{falcke2012_BPJ} and 0.0165-0.0645~nN~\cite{radmacher2011} per
micrometer leading edge. The values of $f\delta_0/k_BT$ resulting from these force
densities are in the range 10-70. That entails $\delta/\delta_0\lesssim 1$ (see
Fig.~\ref{fig5}) in agreement with modelling results in ref.~\cite{falcke2012_BPJ}.

To assess the effects of branching on network properties, we calculate the ratio $R$ of the force exerted by a branched network to the force of an unbranched network with the same angular distribution. We use the same number of filaments in the unbranched network as there are branched structures in the branched network:
\begin{equation}\label{eq:ratio}
  R(\delta) = \frac{<f_b>}
  {\frac{1}{\Delta\omega}  \int_{\omega_{min}}^{\omega_{max}} d\omega \: f_t(\delta_0/\cos{\omega},\delta,\omega)}.
\end{equation}
The ensemble averages of branched and unbranched networks with the uniform orientation distribution between -70$^\circ$ and 70$^\circ$ behave very similar under compression (Fig.~\ref{fig6}, squares). This ratio has the remarkable property of depending only very weakly on $\delta$. It is almost independent of the persistence length $l_p$ and the graft stiffness $K$ as well (data not shown). Hence, the behavior of both networks scales very similar in dependence on these parameters. The ratio is about 2, i.e., the average force per filament tip for a given value of $\delta$ is the same for branched and unbranched networks.

However, there are obvious qualitative differences between single filaments and branched structures illustrated by the non-trivial dependency of the properties of branched structures on the branch point position in Figs.~\ref{fig:graftpoint} and \ref{fig:graftpointomega}. Indeed, if we use a narrow orientation distribution around $\omega=35^\circ$ the value of $R$ increases with increasing compression. Therefore, we also investigate non-uniform measured lamellipodial orientation distributions to investigate whether the dependency of $R$ on the distribution affects the behavior of lamellipodial networks. Distributions peaked either at $\omega=0$ or $\omega=\pm 35^\circ$ have been observed in refs.~\cite{borisy2001,borisy2003,schwarz2012}. The distribution in ref.~\cite{borisy2001}, their Fig.~4, is typical for the measured non-uniform distributions and can be approximated by
\begin{eqnarray}\label{eq:malydist}
 \lefteqn{P(\omega)=0.008012\left[\rm{e}^{\frac{-(\omega-35)^2}{2\cdot 20.5^2}}+\rm{e}^{\frac{-(\omega+35)^2}{2\cdot 20.5^2}}\right]}&&\\ &&+0.0006907,\ \ \  -85^\circ \leq \omega \leq 85^\circ
.\nonumber
\end{eqnarray}
The dots in Fig.~\ref{fig6} show the results for $R$. They are very similar to the results for the uniform distribution. Obviously, the width of 40$^\circ$-45$^\circ$ of the peaks in lamellipodial orientation distributions is too large for an essential effect of branching on the parameter dependencies.

\section{Discussion}

We investigated the properties of branched filaments grafted into an elastic graft. Their stiffness has a maximum in its dependence on the branch point position. Branched structures with the optimal graft point position can be more than four times as stiff as a single filament with the same tilt angle as the mother filament (Fig.~\ref{fig:graftpoint}), while requiring only 1.5 times the polymer length of the mother filament.

The mechanical properties of branched F-actin networks depend on their orientation distribution. With measured lamellipodial distributions, networks of branched structures are about twice as stiff as unbranched networks. The total network force of branched networks scales essentially the same as the one of unbranched networks with the parameters F-actin persistence length, graft stiffness and compression ($\delta$). An intuitive explanation would be, that in the end it is the single filament behaviour determining the stiffness for both unbranched and branched filaments, since the branch leans on the (single) mother filament when experiencing a force.

Our theory considers individual branched filaments and thus establishes the constitutive relations on which complex network studies including cross-linking can be based. We considered F-actin networks as defined by their geometrical property distributions without interactions of filaments by cross-linking or entanglement. This implies that we cannot account for visco-elastic properties. A variety of evidences suggests the existence of a region close to the leading edge, where cross-linking is not dominating the network properties and to which our theory directly applies. Measurements of the ratio of number of the cross-linkers to the number of actin molecules in fibroblasts show the existence of a gradient for $\alpha$-actinin and ABP-280/filamin. The number ratios are low in a region juxtaposed to the leading edge with a depth of about 1.5~\um (see Fig.~5 of ref.~\cite{svitkina99}). Svitkina and Borisy conclude from these results and structural information from electron micrographs that the impact of ABP-280 and $\alpha$-actinin on filament cross-linking is likely to be expressed more deeply in the cytoplasm~\cite{svitkina99}. Their statement is supported by the immediate response of the leading edge of fish keratocytes to small forces indicating also weak cross-linking in the network region close to the leading edge~\cite{falcke2012_BPJ,Radmacher2006,radmacher2011}. This suggests that while understanding of the visco-elastic properties of the network in the lamellipodium bulk requires taking cross-linking into account, our ensemble average is applicable to a network region close to the leading edge. The reproduction of both the weak and strong force responses of the lamellipodium leading edge measured in force-velocity relations by our network calculations strongly supports that conclusion (Fig.~\ref{fig5}).

We did not take contributions from entanglement or excluded volume effects into account when calculating the network forces. This implies that $R$ provides only a meaningful approximation, if these effects are similar in branched and unbranched networks. To the best of our knowledge, that has not been investigated quantitatively yet. We can only provide heuristic considerations in favour of our assumption based on comparing a variety of simulations with and without excluded volume effects with force-velocity measurements.

Model networks of semi-flexible filaments not taking into account excluded volume effects reproduce the elastic properties measured in force-velocity experiments quantitatively~\cite{falcke2012_BPJ}. Schreiber et al. simulated the force-velocity relation of motile cells with rigid rods as model filaments taking excluded volume effects into account~\cite{duke2010}. They found excluded volume effects to be stronger in branched than in unbranched systems. The model network of Schreiber et al. shows a response to external forces in the force-velocity relation at about 8~nN/\um~\cite{duke2010}. However, the lamellipodium leading edge exhibits elastic responses to forces smaller than 0.05~nN/\um in experiments~\cite{falcke2012_BPJ,Radmacher2006,radmacher2011}. Additionally, bending of filaments has been observed in lamellipodia~\cite{Small2008,burnette2011,small2012}, i.e., filaments do not behave like stiff rods. Hence, the lamellipodium network is likely to be in a parameter regime where excluded volume effects are less relevant than suggested by a network of stiff rods.

Branching has also been observed with microtubule~\cite{vale2013}. The branching angle varies between 0$\,^{\circ}$ and 90$\,^{\circ}$, and it is not known, how rigid the connection to the mother filament is. If the branch is rigidly connected, our results should apply also to these branched structures with the adapted persistence length (a few millimeters~\cite{howard93}) and $\gamma$-values.

In summary, single branched filaments with intermediate branch point positions and tilt angles exert larger forces and are stiffer than two unbranched filaments. The stiffness of a whole network of branched filaments is largest, if the orientation distribution is sharply peaked around $\pm\gamma/2=\pm35^\circ$. For lamellipodial orientation distributions, the stiffness of branched and unbranched networks scales approximately the same with a variety of parameters (Fig.~\ref{fig6}), suggesting that the effect of branching on network stiffness can be accounted for by rescaling the filament number of an unbranched network. These results are in agreement with the elastic properties of lamellipodia found in force-velocity measurements with fish keratocytes~\cite{falcke2012_BPJ,Radmacher2006,radmacher2011}.

\section{APPENDIX}
\subsection{Tilted filament}
The Green function of a free semiflexible polymer in two spatial dimensions (see Fig.~\ref{fig:configuration}) and in the weakly bending limit satisfies the partial differential equation (PDE)
\begin{equation}
[\frac{\partial }{\partial s}+\theta \frac{\partial }{\partial y}-\frac{1}{l_p}\frac{\partial^2 }{\partial \theta ^2}]G(s, y,\theta\mid 0, y_{0},\omega)=0.
\end{equation}
The arc length of the filament contour is denoted $s$ here. The boundary condition
\begin{equation}
\lim_{s\rightarrow 0}{G(s, y_{s},\theta_{s}|0, y_{0},\omega)}=\delta (\theta-\omega)\delta (y_{s}-y_{0})
\end{equation}
realizes the graft. The solution is Eq.~\ref{eq:Green1}, after a switch from the coordinates $(s,y,\theta)$ to $(x,y,\theta)$.
We obtain the Green function for a tilted filament simply by a rotation:
\begin{eqnarray}\label{eq:GL_tilted}
 \lefteqn{G_{L}(x,y,\theta|x_0,y_0,\omega) \propto}&&\\ & \mbox{exp}[-\frac{3l_p}{L^3}((y-y_{0})\cos(\omega)-(x-x_{0})\sin(\omega))^2 \nonumber\\&-\frac{l_p}{L}(\theta-\omega)^2]  \nonumber \\
 &\times \mbox{exp}[+\frac{3l_p}{L^2}((y-y_{0})\cos(\omega)-(x-x_{0})\sin(\omega))(\theta-\omega)] \nonumber \\
 &\times \delta [(x-x_{0})\cos(\omega)+(y-y_{0})\sin(\omega)-L].\nonumber
\end{eqnarray}
The probability distribution of the position $x$ of the endpoint follows by integration (see Eq.~\ref{endpoint1})
\begin{equation}
P_{t }(x|x_0)\propto \exp(-\frac{(x-x_0-L\cos(\omega))^2}{\sigma_{t0}^2})
\end{equation}
where
\begin{equation}
\sigma_{t0}^2=\frac{4L^3\sin(\omega)^2}{3l_p}.
\end{equation}
Representing the stiffness of the graft by a spring and averaging over all grafting points, $x_0$, yields:
\begin{equation}
P_t(x)=\frac{1}{N_t}\exp(-\frac{(x-L\cos(\omega))^2}{\sigma_t^2}).
\end{equation}
Here $N_u$ accounts for the proper normalization and
\begin{equation}
\sigma_t^2=\sigma_{t0}^2+\frac{2}{K}.
\end{equation}
The entropic force on the wall is then explicitly given by:
\begin{equation}\label{eq:force of single filament}
f_t(\delta)=\frac{2k_BT}{(\sqrt{\pi} \sigma_t)}\frac{\exp(-\frac{(\delta-L\cos(\omega))^2}{\sigma_t^2})}{\mbox{erfc}(\frac{L\cos(\omega)-\delta}{\sigma_t})}.
\end{equation}

\subsection{Branched filament}

Starting from Eq.~\ref{branched1} and averaging over graft point positions $x_0$, we obtain
\begin{equation}\label{eq:Pb}
P_{b }(x_2,x_3)=(\frac{(\mbox{det(M)})^\frac{1}{2}}{\pi}) \exp(\eta_{i}M_{ij}\eta_{j}),
\end{equation}
where
\begin{equation}
\eta={{x_2-L_1\cos(\omega)-L_b\cos(\omega-\gamma)}\choose{x_3-(L_1+L_3)\cos(\omega)}}
\end{equation}
and $M$ is a $2\times2$-Matrix with components
\begin{equation}
M_{11}=-\frac{l_p}{C}[K(L_1+L_3)^3\sin(\omega)^2+\frac{3}{2}l_p]
\end{equation}
\begin{eqnarray}
M_{12}=M_{21}=&\frac{l_p}{C}[KL_1^2(L_1+\frac{3}{2}L_3)\sin(\omega)^2+\frac{3}{2}l_p+\nonumber\\ &\frac{3}{2}KL_1L_b(L_1+2L_3)\sin(\omega)\sin(\omega-\gamma)]\nonumber
\end{eqnarray}
\begin{eqnarray}
M_{22}=&\frac{-l_p}{C}[KL_b^2(3L_1+L_b)\sin(\omega-\gamma)^2+\frac{3}{2}l_p+\nonumber\\
&3KL_1^2L_b\sin(\omega)\sin(\omega-\gamma)+KL_1^3\sin(\omega)^2],\nonumber
\end{eqnarray}
and
\begin{eqnarray}
\lefteqn{C=K\sin(\omega)^2[(L_1+\frac{4}{3}L_3)L_1^3L_3^2\sin(\omega)^2}&\nonumber\\&-2(L_1^2-2L_3^2)L_1^2L_bL_3\sin(\omega)\sin(\omega-\gamma)]   \nonumber\\&+L_b^2K\sin(\omega)^2\sin(\omega-\gamma)^2[L_1^4+\frac{4}{3}(L_1^3L_b+L_bL_3^3)\nonumber\\&+4L_1^2L_bL_3+4L_1L_3^2(L_b+L_3)]    \nonumber \\
 &+6l_p[(L_1+\frac{1}{3}L_3)L_3^2\sin(\omega)^2-2L_1L_bL_3\sin(\omega)\sin(\omega-\gamma)\nonumber\\&+(L_1+\frac{1}{3}L_b)L_b^2\sin(\omega-\gamma)^2]\nonumber
\end{eqnarray}

\subsection{Parameter values}

The independent parameters of the model are the F-actin persistence length $l_p$, the equilibrium distance of the filament tips from the graft plane $\delta_0$, the stiffness $K$ of the graft, and the branching angle $\gamma$ which is fixed by the Arp2/3 complex at about $70\,^{\circ}$~\cite{pollard98,Volkmann2001,svitkina2011}. The branch point position is uniformly distributed and the tilt angle $\omega$ according to the above distributions. The tilt angle and $\delta_0$ fix the mother filament contour length as $L=\delta_0/\cos(\omega)$.

We explained in the Discussion, that a region juxtaposed to the leading edge membrane with a width of 1.0-1.5~\um is similar to an experimental realization of the network configuration in Fig.~\ref{fig:configuration}, since it is weakly cross-linked. Values for $\delta_0$ suggested by these observations are in the range 1.0-1.5~\umns. Branching will have an effect on network properties in a configuration like shown in Fig.~\ref{fig:configuration}, if most filaments are branched, i.e., if $\delta_0$ is larger than the average branch distance.

The average branch distance in steadily moving cells has been a matter of debate in recent years. Svitkina et al. concluded 20-50~nm from early electron micrographs of lamellipodia from Xenopus keratocytes and fibroblasts~\cite{svitkina99}, and 50-200~nm from another study in fibroblasts~\cite{Bear2002}. Later measurements substantially increased that value. The average branch distance has been determined to be about 800~nm for B16 melanoma cells and fish keratocytes in ref.~\cite{small2012}. Other studies provide number densities of branch points per lamellipodium area for 3T3 fibroblasts. Calculating the average branch distance from that branch point density implies assumptions on the F-actin concentration. The number of filaments per micrometer lamellipodium width in a distance of 0.1-1~\um from the leading edge is 170-190~\cite{small2010}. Taking into account that filament orientation is approximately uniformly distributed between 0$^{\circ}$ and 60$^{\circ}$~\cite{Small2008}, this density means 1.16(170-190)~\um filament contour length per \umsqu lamellipodium area. The factor 1.16 arises from averaging over all tilt angles of the filaments. Yang and Svitkina measured 277 branch points/\umsqu~\cite{svitkina2011} in the same sample, i.e. an average branch point distance in terms of contour length between 700~nm and 800~nm. Small et al. measured less than 225 branch points/\umsqu~\cite{small2011}, i.e. an average distance of more than 860~nm.

The {\it in vivo} persistence length is not known. We can only conclude a reasonable range from {\it in vitro} measurements. Results from fluctuation analysis and measurements of network elastic properties yield values of the {\it in vitro} persistence length of 15-18~\um~\cite{howard93,frey2002,gardel2004,carlier95}. The filaments in these experiments were stabilized with phaloidin which most likely increases the value of $l_p$. Filaments labeled with rhodamine but not stabilized with phaloidin exhibited values of $l_p$ between 9~\um and 13.5~\um \cite{carlier95}. The {\it in vivo} persistence length might be even shorter, since cofilin can substantially reduce it even down to 2.2~\um~\cite{delacruz2008,delacruz2010}.

\section*{ACKNOWLEDGMENTS}

AZ acknowledges financial support by the DFG through Grant No. SFB 937/A1. MF acknowledges financial support by the DFG through Grant No. FA350/11-1.

\section*{References}
\bibliographystyle{unsrt}
\bibliography{testbib}

\end{document}